# Engineering thermal emission with enhanced emissivity and quality factor using bound states in the continuum and electromagnetically-induced absorption

**Authors**
Guodong Zhu [1,2], Ikjun Hong [1,2], Theodore Anyika [1,2], Maxwell T. Ugwu [1,4], J. Ryan Nolen [3], Mingze He [5], Joshua D. Caldwell [1,3,4], Justus C. Ndukaife [1,2,3,4]*

**Affiliations**
[1] Vanderbilt Institute of Nanoscale Science and Engineering, Vanderbilt University, Nashville, TN 37235, USA.
[2] Department of Electrical and Computer Engineering, Vanderbilt University, Nashville, TN 37235, USA.
[3] Department of Mechanical Engineering, Vanderbilt University, Nashville, TN 37235, USA.
[4] Interdisciplinary Materials Science and Engineering, Vanderbilt University, Nashville, TN 37235, USA.
[5] Photonics Initiative, Advanced Science Research Center, City University of New York, New York, NY 10031, USA

**Abstract**

Metal-based thermal metasurfaces exhibit stable spectral characteristics under temperature fluctuations, in contrast to more traditional gray- and near black-bodies, as well as some dielectric metasurfaces, whose emission spectra shift with changing temperatures. However, they often suffer from limited quality (Q) factors due to significant non-radiative ohmic losses. In this study, we address the challenge of achieving high emissivity and Q-factors in metal-based thermal emitters. By leveraging the coupling between a magnetic dipole resonance and two bound-state-in-continuum (BIC) resonances to achieve electromagnetically induced absorption (EIA) in an asymmetric metallic ring structure, we design a metal-based thermal metasurface with a near-unity emissivity (0.96) and a Q factor as high as 320 per simulations. Experimental validation yields an emissivity of 0.82 and a Q factor of 202, representing an approximately five-fold improvement in the experimentally measured Q factor compared to the state-of-the-art metal-based thermal metasurfaces. Our work offers a promising approach for developing efficient, narrow-band, directional thermal emitters with stable emission spectra across a wide temperature range.

**Introduction**

Thermal emission is considered one of the most important approaches to produce light at infrared (IR) wavelengths (*1*). However, typical thermal emission is broadband, omnidirectional, and unpolarized owing to the spontaneous, incoherent nature of the process (*2*). To achieve coherent light emission from thermal radiation, various artificial nanostructures and physical mechanisms have been utilized to tailor the Q factor, polarization, and directionality. In 2002, Greffet et al. (*3*) pioneered the development of a spatially coherent thermal emitter using a silicon carbide grating that supports surface phonon polaritons (SPhPs) that could launch and directionally emitted using heat. This remarkable effect is a result of engineering surface waves, which leads to coherent emission in the far field. Later works from Schuller et al. demonstrated the ability of anisotropic structured materials (here SiC microwires) could support polarized thermal emission (*4*), while Wang et al. demonstrated that local antenna structures could be used to induce narrowband spectrally coherent light (*5*). Subsequently, Ikeda et al. (*6*) demonstrated the control of thermal emission using metal gratings, supporting surface plasmon polaritons (SPPs). However, SPPs and SPhPs suffer from relatively large propagation losses, limiting their temporal coherence.

Towards overcoming the challenges of metal ohmic losses in polaritonics, researchers have explored alternative materials and structures to support narrow-band resonances. Some of them include photonic crystals (*7*, *8*), metasurfaces (*9*, *10*), and films stacks (*11*, *12*). In recent years, metallic and dielectric thermal metasurfaces that support bound states in the continuum (BICs) have emerged as a promising platform offering the capability for narrow-band emission, wavefront engineering, and flexible design possibilities. Originally predicted in early theoretical works in quantum mechanics, BICs theoretically exhibit infinite radiation Q factors and thus, vanishing linewidths (*13–21*). By breaking symmetry, adjusting the incident angle, and/or tuning the metasurface parameters, the linewidth of BIC resonances can be finely tuned for practical applications. Recently, Overvig and Nolen *et al.* (*2*, *22*) demonstrated that the polarization and emission directionality can be controlled using elliptical structures with a non-local quasi-BIC resonance and local geometric phase. Yang *et al.* (*23*) developed a slotted elliptical metallic thermal metasurface supporting a quasi-BIC resonance, capable of emitting narrow-band photons, such that the emission spectrum peak position remains stable across a wide range of temperatures. Similarly, Sun *et al.* (*24*) showed that quasi-BIC dielectric structures, leveraging flat-band engineering, can enhance the Q factor up to approximately 230. However, their emission spectral peak position varied with temperature and exhibited reduced spatial coherence. This spectral drift is attributed to the strong dependence of the refractive index of dielectric materials (e.g., silicon and germanium) on temperature (*25–28*), owing to the thermo-optic effect.

In contrast, thermal metasurfaces based on metallic structures generally tend to have more stable emission spectral peak positions in the mid- to long-wave infrared ranges (*23*). This stability arises because the imaginary part of the refractive index often dominates in metals at these wavelengths, resulting in a photon penetration depth much smaller than the emission wavelength. Consequently, the resonance frequency is determined primarily by geometry (*23*, *29*, *30*). Nonetheless, achieving near-unity emissivity in a narrow band using metallic structures still hinges on meeting the critical coupling condition, which involves carefully balancing radiative and non-radiative losses (*31*, *32*). There is typically a trade-off between a high Q factor and high emissivity, and non-radiative losses affected by both material properties and fabrication imperfections can be difficult to minimize. As a result, practical Q factors and emissivities remain limited; to date, the highest experimentally measured Q factor for metal-based thermal emitters is 42 (*23*).

In this work, we present an approach where we leverage the coupling of multiple resonances as a means to achieve near-unity emissivity and Q factors exceeding 200, which is well above the typical reported range of 10 to 60 in metal-based thermal metasurfaces. We show that by precisely tuning the coupling between two BIC resonances and a magnetic dipole resonance, metallic metasurfaces can in principle be used to achieve both a high Q factor (320) and near-unity emissivity (0.96), simultaneously. We attribute the absorption enhancement to the classical analog of electromagnetically induced absorption (EIA). In quantum systems, EIA arises from constructive quantum interference of excitation probabilities at different energy levels, leading to the enhancement of absorption (*32–34*). As to the physical implementation, we employ a metal-insulator-metal (MIM) configuration to prevent transmission through the structure, while the aluminum oxide spacer provides a means for tuning the coupling strength of the resonances as depicted in Fig. 1. This design experimentally exhibits Q-factors (202) and emissivities (0.82) in good agreement with the upper limits predicted by simulations.

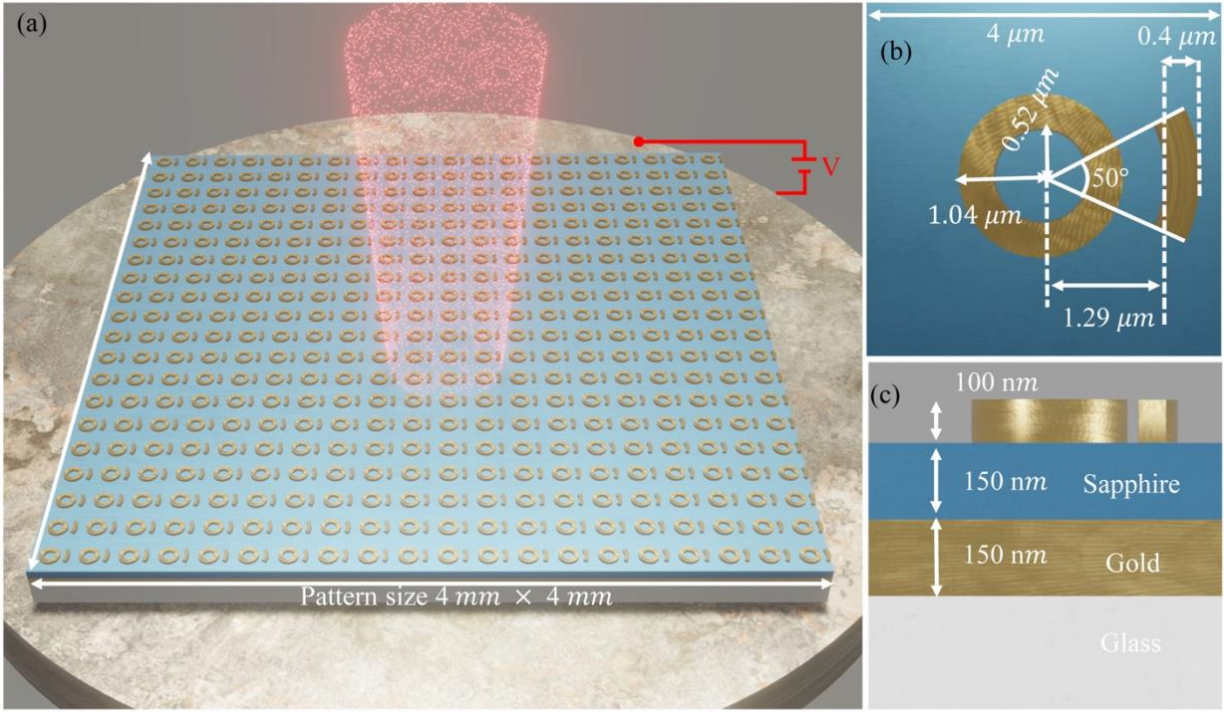

**Fig. 1. Schematic of the metal-insulator-metal (MIM) configuration used in the study.** (a) shows the schematic of the thermal metasurface for narrowband directional thermal emission. (b) shows the top view of a unit cell of the metasurface comprising a complete circular gold ring and a segmented gold ring. (c) shows the side view of a unit cell. The gold reflector is used for preventing optical transmission, while an aluminum oxide spacer layer tunes the coupling of the resonances induced in the thermal metasurfaces to optimize the emission.

## Results

According to Kirchhoff's Law, achieving a perfect emitter is inherently equivalent to creating a perfect absorber. Thus, we focus on the absorption response of our metasurface in simulations. Through numerical simulations, we achieved an ultra-narrowband thermal emission spectrum near the mid-infrared wavenumber of 2450 $cm^{-1}$ by carefully engineering the coupling between three resonances. The simulation results in Fig. 2(a) show Q factors of approximately 320 and near-unity emissivities of 0.96 for X polarization, while experimental validation demonstrates Q factors of 202 and emissivities of 0.82 in Fig. 4(a). The peak position of the thermal emission spectrum mediated by the coupled resonances is tunable by adjusting the periodicity and radius of the metasurface ring elements while maintaining high Q factors and emissivity (see Supplementary Section S3). Two of the three coupled resonances, Resonance 1 (M1) and Resonance 3 (M3), are associated with bound states in the continuum (BICs). Resonance 1 (M1) is a dual-quasi-BIC resonance that combines two types of BICs: the symmetry-protected BIC (SP-BIC) (*14*) and the Friedrich–Wintgen BIC (FW-BIC) (*13*), while resonance 3 (M3) is a symmetry-protected BIC (SP-BIC) (confirmed in Supplementary Material Section S5). Both M1 and M3 are dark states with theoretically infinite radiative Q factors and are weakly coupled to free space. Thus, they exhibit vanishing spectrum flanking Resonance 2 (M2) on the left and right in Fig. 2(a), which is characteristic of BICs. For the center resonance M2, we find that it is dominated by an in-plane magnetic dipole mode (confirmed by multipolar decomposition analysis (*35*), in Fig. S6), acting as a bright state that can directly couple to free space. Due to the intrinsically high Q factor of the two involved BIC modes, the coupled resonance demonstrates an ultra-narrow linewidth in Fig. 2(a). The electric field distribution for the three resonances is depicted in Fig. 2(b), where it is evident that M2 is a magnetic dipole mode pointing in the in-plane direction. As for the near-unity emissivity, it is achieved through the coupling of the resonances that drive the system towards the critical coupling condition, thereby enhancing the absorption. This phenomenon can also be termed as a classical analog of EIA. The coupling process of the three resonances is illustrated in the four-level system for the metasurface in Fig. 2(c). Only M2 can be directly excited from free-space with relatively larger radiative loss $\gamma_2$, while the M1 and M3 are excited through their near-field coupling with M2 (*33*).

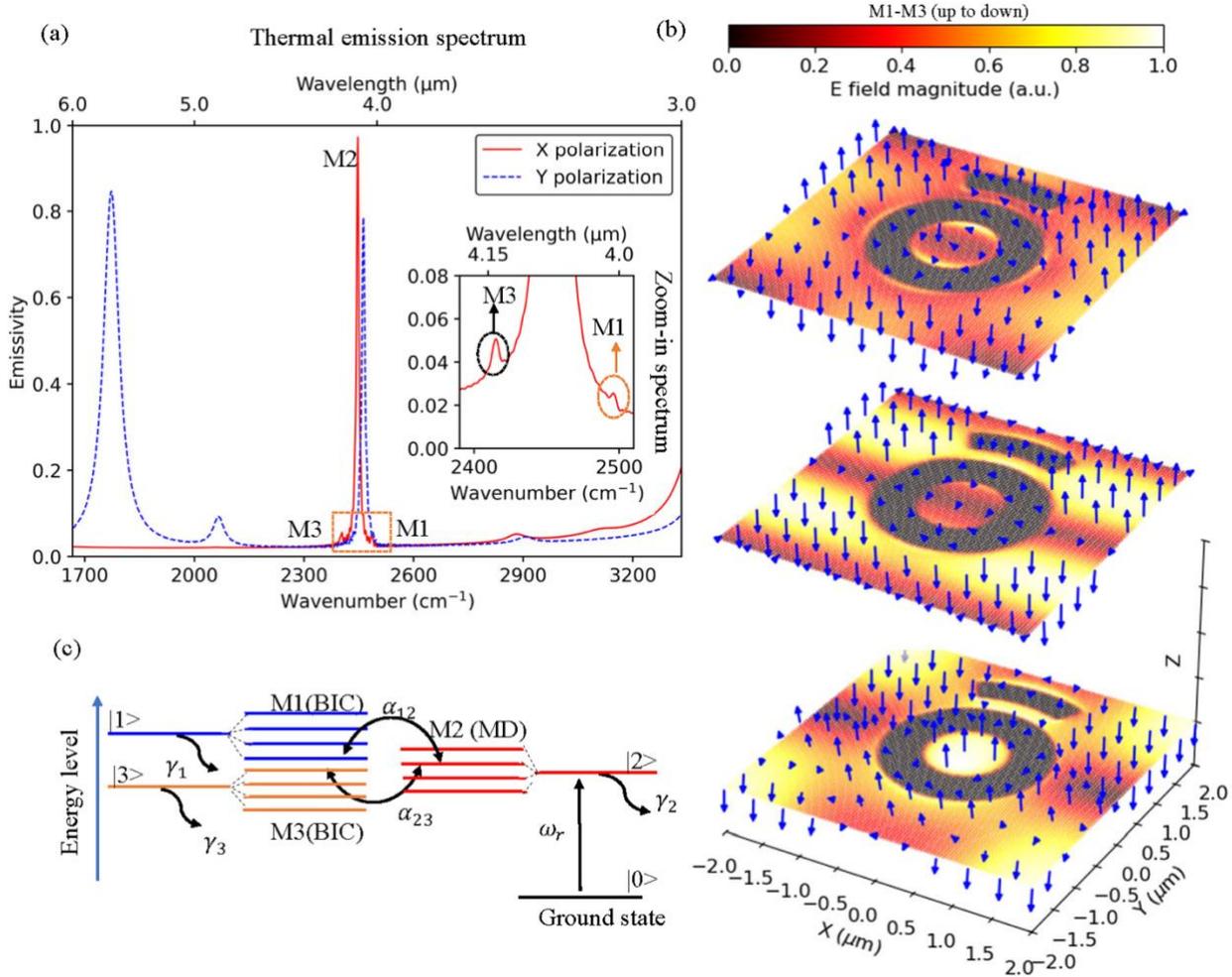

**Fig. 2. Simulated emission spectrum showing a high Q factor and near-unity emissivity, along with the electric field distribution of the three coupled resonances.** (a) shows the emission spectrum for X and Y polarization. The spectrum for X polarization indicates three resonances around 2450 cm$^{-1}$: Resonance 1 and Resonance 3 are BIC modes with vanishing spectral peaks, while Resonance 2 is a resonance mode in the center with strong absorption. By tuning the coupling between the three resonances, a high Q factor and a near-unity emissivity can be achieved. Inset shows a zoomed-in spectrum for X polarization. (b) shows the field distribution of the three resonances. Resonance 1 (M1) is a dual-quasi-BIC combining symmetry-protected BIC (SP-BIC) and Friedrich-Wintgen BIC (FW-BIC) with theoretically infinite Q factor, Resonance 2 (M2) is an in-plane magnetic dipole mode which can be excited directly from space, Resonance 3 (M3) is another SP-BIC mode with theoretically infinite Q factor. Because of their weak coupling to free-space, M1 and M3 can be termed as "dark" states, in the contrast, M2 can be termed as a "bright" state. (c) coupled four level systems for the metasurface. Only M2 can be excited from free space, while the M1 and M3 can be excited through their coupling with M2. $\gamma_i$ represents the radiative loss, $\alpha_{ij}$ are the coupling constant between the resonance i and resonance j.

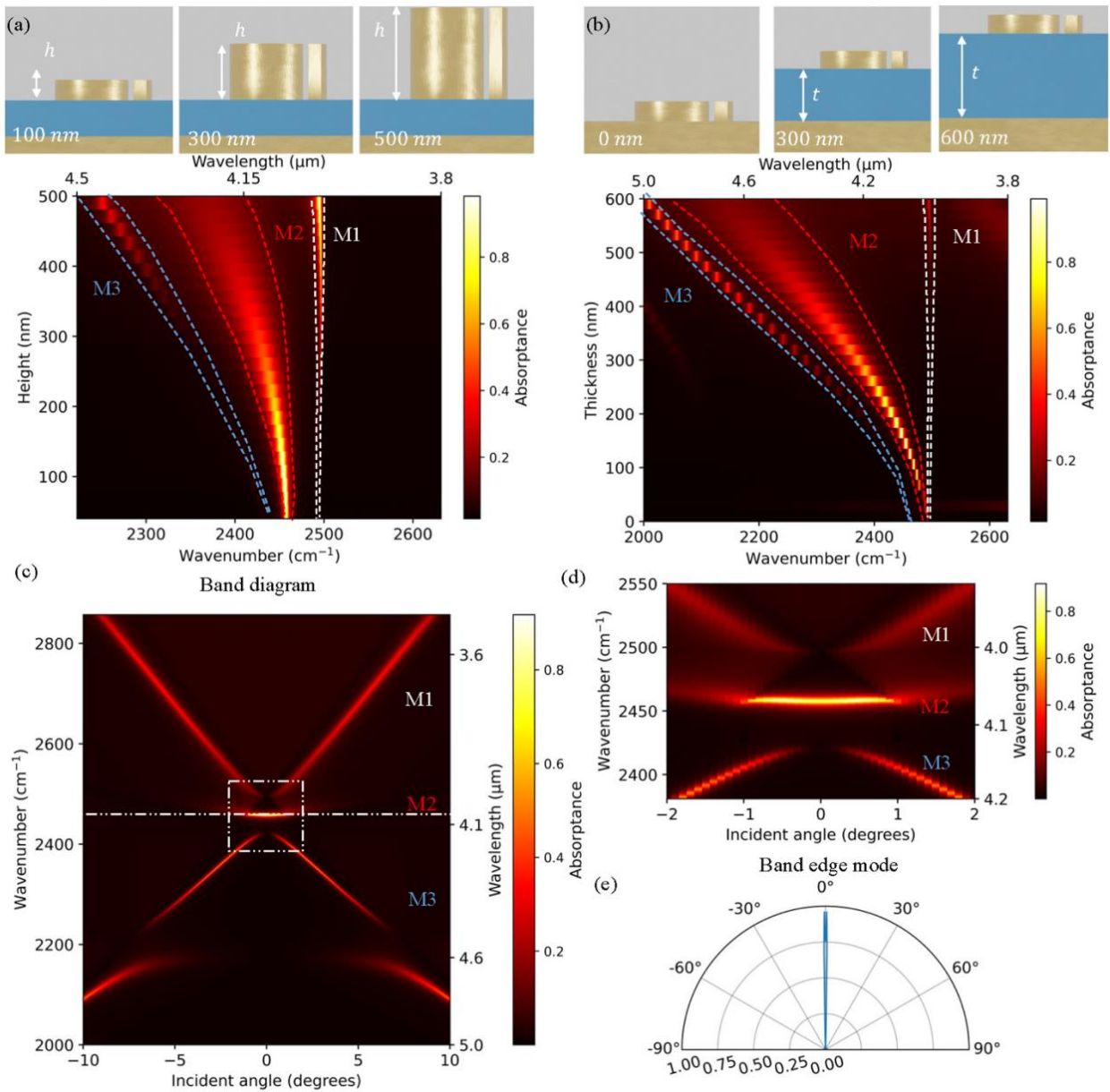

**Fig. 3. Mode coupling demonstration from parameter tuning and band diagram perspective** (a) shows structure's absorptance as a function of pillar height when the spacer layer thickness is fixed at 150 nm. (b) shows structure's absorbance as a function of spacer layer thickness when the pillar height is 100 nm. The absorbance reaches approximately 0.9 when the spacer layer is 150 nm thick. The absorptance and Q factor are tunable by adjusting the spacer layer thickness and pillar height. When the three resonances get close, the strong coupling between enhances the Q factor and absorbance. (c) shows the band structure of the three modes in momentum space. M1, M2, and M3 are strongly coupled near the Γ points (within 1˚), significantly enhancing the absorptance of Mode 2 up to 0.9. (d) is the zoom-in figure from the white box of (a). (e) The polar diagram of the band-edge mode (white line in (a)) further confirms the good directionality of the emitter.

Before diving into the full explanations of the detailed physics, a simplified "balls and sticks" model of the structural design can provide an intuitive understanding of the underlying mechanism. High-symmetry circular disk structures typically support BIC modes (*2*). In theory, ideal BICs exhibit infinite lifetimes or Q factors (vanishingly small resonance linewidths), rendering them invisible in the spectrum. However, through perturbative approaches, such as engineering the material losses or introducing symmetry mismatches, ideal BICs can be transformed into quasi-BICs. Here, we break the in-plane symmetry of a gold disk by adding a gold segment ring on its side, enabling the structure to support both SP-BIC and FW-BIC simultaneously. This dual-quasi-BIC resonance enhances the Q factor in plasmonic structures (*36*). Nevertheless, this design also introduces multiple unwanted resonances near the BIC resonances in the presence of a spacer and reflector layer. To eliminate these undesired resonances, we transform the disk structure into a ring structure and adjust the position of the gold segment ring away from the ring. As a result, only two BIC resonances and a magnetic dipole dominate resonance remain in the target wavelength range. Finally, by optimizing the coupling between these resonances, we achieved a single ultra-narrow absorption peak in the spectrum.

To numerically demonstrate how the Q factor and emissivity is enhanced due to the coupling of the two BIC resonances with the bright state, we constructed parameter sweep diagrams by adjusting the spacer layer thickness and gold structure height (as depicted in Fig. 3(a)(b)). In the first scenario, we fixed the spacer layer thickness at 150 nm, adjusting the height of the gold ring structure from 500 nm to 50 nm, in 10 nm steps. The results show that the resonance frequencies of M1 and M3 are approaching that of M2, while the absorption of M2 is increasing and the linewidth becomes narrower (Fig. 3(a)). Then, we fixed the height of the gold ring structure at 100 nm, tuning the thickness of the spacer layer from 600 nm to 0 nm, and similar trends were observed where the Q factor and emissivity are boosted when the three resonances become coupled (as depicted in Fig. 3(b)). To further elucidate the underlying mechanism of our absorption enhancement mathematically, we applied temporal coupled mode theory (TCMT) to derive the system absorption expression under the three resonance coupling scenario (Materials and Methods section). The expression shows that near-unity absorption can be realized by tuning the coupling parameter $\alpha_{12}$ and $\alpha_{23}$ between M1/M2 and M3. We also plot the band diagram of the three resonances in momentum space in Fig. 3(c). M1, M2, and M3 are coupled near the vicinity of the $\Gamma$ point (within 1°) and the absorptance of the coupled resonances become significantly enhanced (Fig. 3(d)). Outside this angular range, the absorptance goes down dramatically. Due to reciprocity, this indicates a very good directionality and spatial coherence of our thermal emitter. As shown in Fig. 3(e), the polar diagram of the band-edge mode (white line in Fig. 3(c)) further confirmed this.

To validate our theoretical model and design, we fabricated and experimentally characterized the emission properties of the metasurfaces with an area spanning 4 mm by 4 mm, comprising approximately 1 million unit cells. These structures were fabricated on a glass substrate coated with 150 nm gold reflecting layer and 150 nm aluminum oxide spacer layer. The fabrication was achieved using electron beam lithography and physical vapor deposition. The scanning electron microscope (SEM) image of a fabricated sample is depicted in Fig. 4 (c). The emission spectrum was measured using a Fourier transform infrared (FTIR) spectrometer (Bruker Vertex 70v) equipped with a mercury cadmium telluride (MCT) detector. More details of the fabrication and measurement procedures are provided in the supplementary material section S1. We heated the sample to a temperature of 300 °C and acquired the emission signals. The results depicted in Fig. 4 (a) show a single resonance with a high Q factor of 202 and a peak emissivity of 0.82 for X polarization. The Q factor is calculated with the overall damping rate and the resonance frequency, which are obtained from fitting the spectrum with the "CFTool" toolbox in MATLAB (see details in Materials and Methods section). Compared with the simulation results in Fig.2 (a), the features of the measured spectra agree well with the simulation. For Y polarization, the measurement also matches the simulation, displaying both a sharp and a broad peak within the wavelength range of interest, as depicted in Fig. 4 (b). We note that the increased noise around 2350 $cm^{-1}$ is due to carbon dioxide absorption. The discrepancies between experimental results and simulations can be attributed to several factors including fabrication imperfections, such as surface roughness and stitching errors during electron beam lithography, the finite size of our experimental pattern, and the illumination conditions for simulation versus emission collection during experiments. In simulation, normal incident illumination is assumed, whereas in the experiments, the emitted photons are collected within an angular resolution of approximately 1° using an in-house setup. The dispersive nature of the band structure, as shown in Fig. 3(c), indicates that the experimental results will be affected under off-normal photon collection (see Supplementary Material S4). To enhance the thermal metasurface performance, we can address these issues by fabricating larger samples with more unit cells, optimizing gold deposition, and refining the experimental setup to limit the photon collection angle.

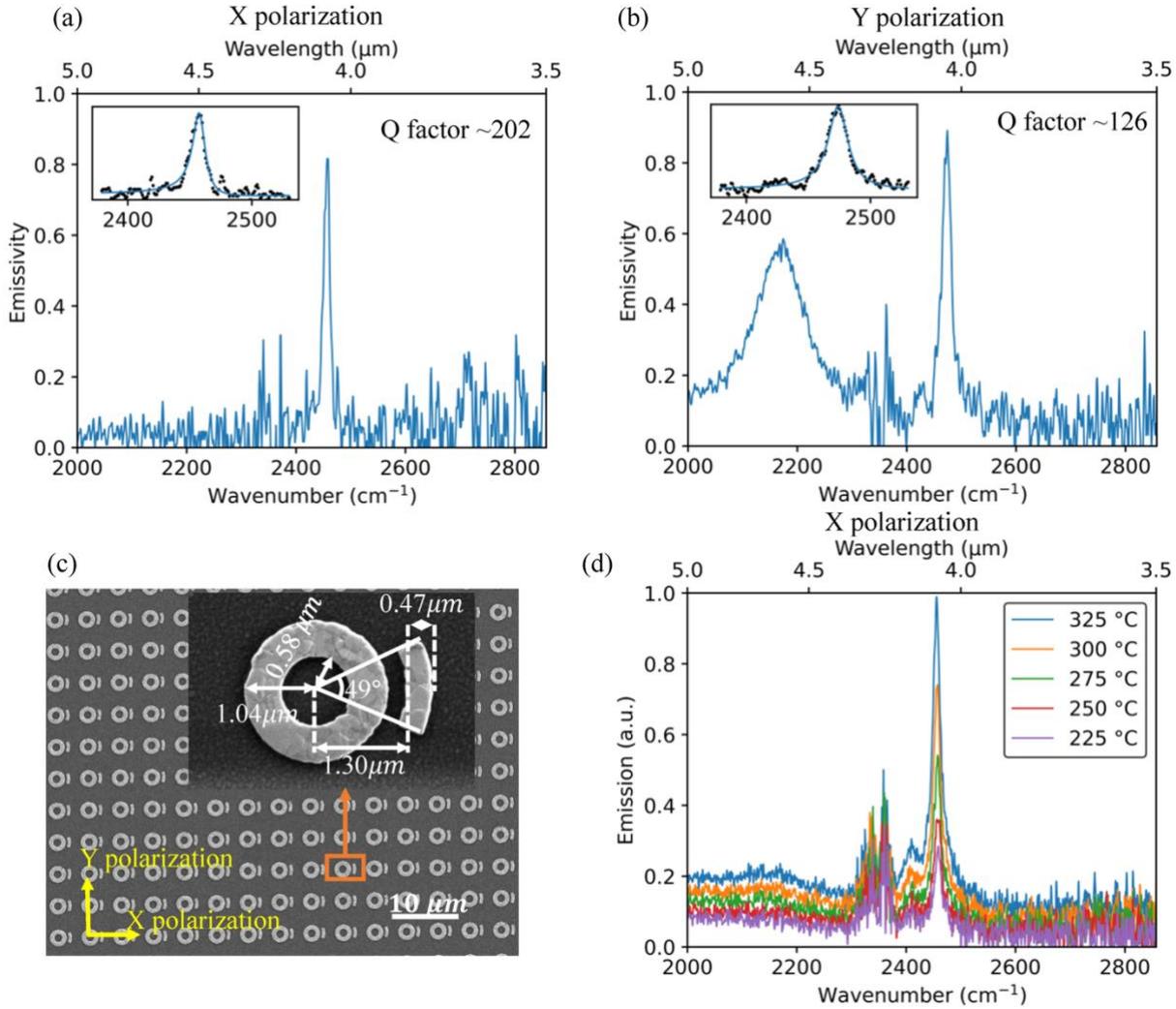

**Fig. 4. SEM of fabricated sample and measured results**. (a) and (b) shows thermal emission spectrum at normal direction when the sample is heated up to 300 °C at X and Y polarization, respectively. The emissivity is obtained from normalizing the emission to the emission of block body at the same temperature. The Q factor is extracted from "CFTool" in MATALAB. Insets The insets show the fitted curve. (c) shows the SEM images of the fabricated sample of 4 mm by 4 mm. (d) The emission spectrum under various temperatures showing the center wavelength exhibits a small shift, indicating temperature insensitivity of the design. The emission goes down when the sample is heated at lower temperature. The emission is normalized by the emission at 325 °C.

We also show that the spectral emission of our metal-based thermal metasurface is resilient to temperature fluctuations. We heated the thermal metasurface and collected the emission spectrum over temperatures ranging from 225 °C to 325 °C during the experiments. As depicted in Fig. 4(d), the resonance wavelength exhibits no noticeable shift over the 100 °C temperature variation range, which indicates a strong robustness to temperature variation for our metal-based thermal metasurface. This represents an important advantage over dielectric-material-based Mie thermal metasurfaces, whereby the emission spectrum shifts with changes in temperature due to lattice expansion (*24*). Overall, thanks to the optimal light-matter interaction induced by the three resonance coupling, our thermal metasurface combines the desirable performance metrics of high

Q, near-unity emissivity, and robustness to temperature variations, which have not been achieved in prior works for metal-based thermal metasurfaces. To be more specific, we have compared our design with several prior works on thermal emitters based on different mechanisms in Table 1.

**Table 1. Comparison between Different Experimentally Measured Thermal Emitter**

| Mechanism | Temperature (K) | Q factor | Emissivity | Resilience against temperature variations | Spectrum tunability | Ref |
|---|---|---|---|---|---|---|
| SPhP | 539 | 210 | 0.60 | NO | Low | Nano Lett. 2021 (*10*) |
| SPP | 573 | 16 | ~1 | YES | High | Nano Lett. 2020 (*37*) |
| TPP | 373 | 775 | 0.3 | NO | High | ACS Photo. 2020 (*38*) |
| BIC | 573 | 42 | 0.56 | YES | High | Optica 2024 (*23*) |
| Band folding | 549 | 224 | 0.5 | NO | High | Nature Comm. 2024 (*24*) |
| EIA with BICs | 573 | 202 | 0.8 | YES | High | This work |

We further show that our thermal metasurface can be optimized to produce consistent spectral emission for varying polarizations by employing a symmetric ring structure, as depicted in Fig. 5(a). Removing the segmented ring transforms the two BIC resonances into fully dark states—undetectable in the emissivity spectrum, while preserving both high Q factor and strong absorption. Fabrication and measurement of the emission achieved by this design align well with the simulation data as depicted in Fig. 5(b), demonstrating the robustness of this EIA-based approach.

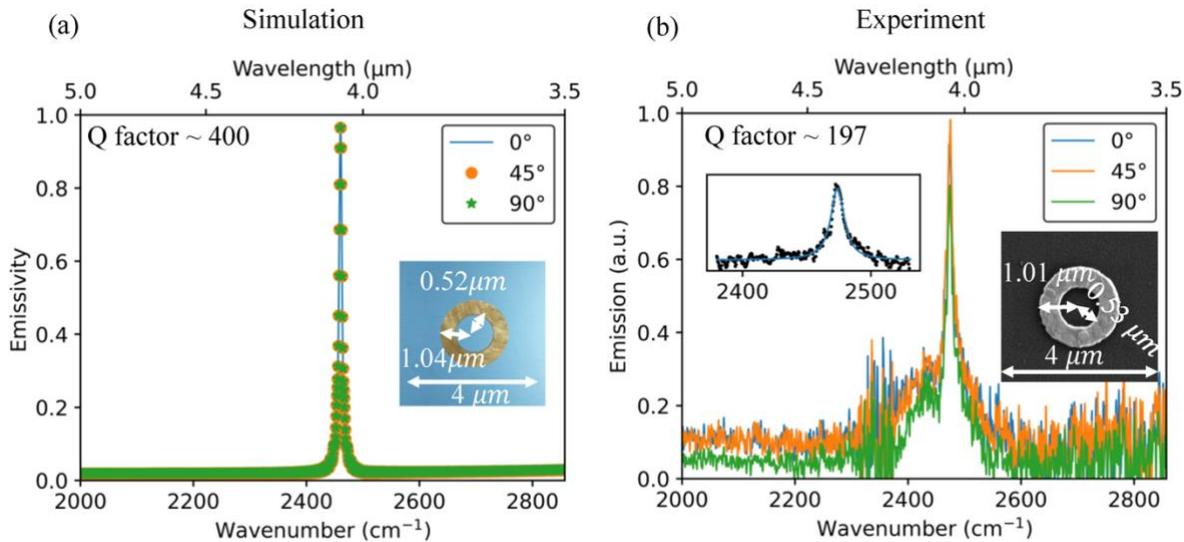

**Fig. 5. Symmetry ring structure for polarization independent emission**. The symmetry ring structure still shows a high Q factor and a high emissivity assisted by EIA. (a) shows the simulation results under the three polarization conditions and (b) show the experimentally measured results for the same polarization conditions as the simulation data. The emission is normalized to the emission at 0 degree polarization. The Q factor are extracted from "CFTool" in MATLAB. The insets show the fitted curve and the SEM image of the fabricated sample.

## Discussion

In this study, we leveraged the coupling between two bound-state-in-the-continuum (BIC) resonances (dark states) and a magnetic dipole mode (bright state) to achieve a high Q factor and

Page **8** of 14

near-unity emissivity in metal-based thermal emitters. By finely tuning this coupling, we realized a classical analog of electromagnetically induced absorption (EIA), which strengthens light–matter interactions and enhances the absorptivity and hence emissivity. Our experimental results confirm these simulation results, demonstrating a single high Q resonance (up to 202) and a peak emissivity of 0.82 at 300 °C for X polarization. Notably, these thermal emission spectra remain stable across a wide temperature range, distinguishing them from previously reported dielectric-based emitters. Importantly, the thermal emission spectral peak position can be readily tuned over a broad range of wavelengths in the mid-IR by scaling the thermal metasurface unit cell elements. Our findings have broad implications for applications requiring narrowband thermal emission, including free-space communication, molecular sensing, medical diagnostics, environmental monitoring, and thermal management (*1*, *2*, *35–39*).

## Materials and Methods

**Materials and Fabrication.** The thermal metasurfaces were fabricated using a standard nanofabrication flow in a cleanroom (shown in the supplementary material Fig. S2.). We deposited Chromium on the top of a clean glass substrate and then deposited gold reflection and aluminum oxide spacer layer. The gold ring structures above the spacer layer were patterned by electron beam lithography with positive photo resist (PMMA). A thin layer (3 nm) of Titanium was used as the adhesion layer. More details of the fabrication flow are provided in the Supplementary materials.

**Simulations.** The simulated absorption spectra of the metasurface under the normal illumination condition were obtained from Tidy3D FDTD. The field distributions were then extracted from the Ansys Lumerical FDTD solution 2024 at the resonant frequencies. We also performed the angular responses simulation of the metsurface in CST STUDIO SUITE 2018. We compared the simulation results obtained from all software and confirmed that the differences between them are small enough. In all the simulations, the refractive indices values of the spacer ($Al_2O_3$) and glass ($Si_2O_3$) substrate were set to 1.685 and 1.46, respectively. The refractive value of gold was taken from Olmon et al.'s data (*39*). The environment for the simulated system was considered as air. For simulations in Tidy3D FDTD, periodic boundary conditions were applied in both x and y directions, and the PML boundary conditions along z direction. We used 64 layers for the upper PML to achieve better absorption at the boundary. Override mesh regions enclosed the resonators, spacer and reflector layers with a maximum mesh step of 10 nm in x, y and z directions. We used the same boundary conditions and mesh for simulations in Lumerical FDTD with Tidy3D. For the simulations in CST STUDIO, the unit cell boundary conditions were set in both x and y directions, and the open boundary condition was set in the z direction. The number of Floquet modes for the wave ports was set as 15 to avoid fake peaks. The Maximum mesh step width was 0.1 μm for the gold structure.

**Temporal coupled mode theory (TCMT) for single resonance.** The dynamic equation (1) and (2) describe the optical process in complex resonant structures (*40–44*). The term *A* represents the modal amplitude, $\tau_r$ is the lifetime of radiative radiation, $\tau_a$ is the lifetime of non-radiative loss. $\omega_r$ is the resonance frequency. $S_-$ and $S_+$ represent the out-going wave and in-coming wave. $\kappa$ is the coupling coefficient of in-coming wave for modal amplitude A. d is the coupling parameter that demonstrates the contribution of modal resonances to the out-going wave. Because of energy conservation and time-reversal symmetry of Maxwell's equations, we have $\kappa = d = \sqrt{2/\tau_r}$. The term C depicts the background scattering matrix. From equation (4), we can see the maximum absorption $\Lambda = 1$, occurring at $\tau_r = \tau_a$ (critical coupling condition)

$$\frac{dA}{dt} = -i\omega_r A - \left(\frac{1}{\tau_r} + \frac{1}{\tau_a}\right)A + \kappa S_+ \tag{1}$$

$$S_- = CS_+ + dA \tag{2}$$

$$R = \frac{(\omega_r - \omega)^2 + \left(\frac{1}{\tau_r} - \frac{1}{\tau_a}\right)^2}{(\omega_r - \omega)^2 + \left(\frac{1}{\tau_r} + \frac{1}{\tau_a}\right)^2} \tag{3}$$

$$\Lambda = \frac{\frac{4}{\tau_r \tau_a}}{(\omega_r - \omega)^2 + \left(\frac{1}{\tau_r} + \frac{1}{\tau_a}\right)^2} \tag{4}$$

**Temporal coupled mode theory (TCMT) for three coupled resonances.**

We use $\alpha_{ij}$ to represent the coupling constant between the modes. The dynamic equations for the three coupled resonances are:

$$\frac{dA_1}{dt} = -i\omega_1 A_1 - \left(\frac{1}{\tau_{r1}} + \frac{1}{\tau_a}\right)A_1 - i\alpha_{12}A_2 + \kappa_1 S_+ \tag{5}$$

$$\frac{dA_2}{dt} = -i\omega_2 A_2 - \left(\frac{1}{\tau_{r2}} + \frac{1}{\tau_a}\right)A_2 - i\alpha_{12}A_1 - i\alpha_{23}A_3 + \kappa_2 S_+ \tag{6}$$

$$\frac{dA_3}{dt} = -i\omega_3 A_3 - \left(\frac{1}{\tau_{r3}} + \frac{1}{\tau_a}\right)A_3 - i\alpha_{23}A_2 + \kappa_3 S_+ \tag{7}$$

$$S_- = CS_+ + d_1 A_1 + d_2 A_2 + d_3 A_3 \tag{8}$$

Apply the harmonics wave expression, $A_j = e^{-i\omega_j t}$, we have frequency domain expression:

$$i(\omega_1 - \omega)A_1 + \left(\frac{1}{\tau_{r1}} + \frac{1}{\tau_a}\right)A_1 + i\alpha_{12}A_2 = \kappa_1 S_+ \tag{9}$$

$$i(\omega_2 - \omega)A_2 + \left(\frac{1}{\tau_{r2}} + \frac{1}{\tau_a}\right)A_2 + i\alpha_{12}A_1 + i\alpha_{23}A_3 = \kappa_2 S_+ \tag{10}$$

$$i(\omega_3 - \omega)A_3 + \left(\frac{1}{\tau_{r3}} + \frac{1}{\tau_a}\right)A_3 + i\alpha_{23}A_2 = \kappa_3 S_+ \tag{11}$$

As the mode 1 and mode 3 are dark state, the out-coupling parameter $\kappa_1$, $\kappa_3 \approx 0$ and $\frac{1}{\tau_{r1}}, \frac{1}{\tau_{r3}}$ can be neglected as the lifetime of BIC mode is very long, resulting in near-zero values of $\frac{1}{\tau_{r1}}, \frac{1}{\tau_{r3}}$, compared to $\frac{1}{\tau_a}$. Here, we also assume that $\omega_1 \approx \omega_3 \approx \omega_2 = \omega_r$, as they are very close. Solving $A_1$, $A_2$, $A_3$ from equation (9) – (11) and substituting them to equation (8), we can get the reflectance and absorption:

$$R = \frac{\left(\frac{1}{\tau_{r2}} - \frac{1}{\tau_a} - \frac{(\alpha_{12}^2 + \alpha_{23}^2)\frac{1}{\tau_a}}{(\omega_r - \omega)^2 + \frac{1}{\tau_a^2}}\right)^2 + (\omega_r - \omega)^2\left(1 - \frac{(\alpha_{12}^2 + \alpha_{23}^2)}{(\omega_r - \omega)^2 + \frac{1}{\tau_a^2}}\right)^2}{(\omega_r - \omega)^2\left(1 - \frac{(\alpha_{12}^2 + \alpha_{23}^2)}{(\omega_r - \omega)^2 + \frac{1}{\tau_a^2}}\right)^2 + \left(\frac{1}{\tau_{r2}} + \frac{1}{\tau_a} + \frac{(\alpha_{12}^2 + \alpha_{23}^2)\frac{1}{\tau_a}}{(\omega_r - \omega)^2 + \frac{1}{\tau_a^2}}\right)^2} \tag{12}$$

$$\Lambda = \frac{4\left(\frac{1}{\tau_a} + \frac{(\alpha_{12}^2 + \alpha_{23}^2)\frac{1}{\tau_a}}{(\omega_r - \omega)^2 + \frac{1}{\tau_a^2}}\right)\frac{1}{\tau_{r2}}}{(\omega_r - \omega)^2\left(1 - \frac{(\alpha_{12}^2 + \alpha_{23}^2)}{(\omega_r - \omega)^2 + \frac{1}{\tau_a^2}}\right)^2 + \left(\frac{1}{\tau_{r2}} + \frac{1}{\tau_a} + \frac{(\alpha_{12}^2 + \alpha_{23}^2)\frac{1}{\tau_a}}{(\omega_r - \omega)^2 + \frac{1}{\tau_a^2}}\right)^2} \tag{13}$$

From the equation (13), at the resonance $\omega_r = \omega$, to get $\Lambda = 1$ (meet critical coupling condition), we need to let $\frac{1}{\tau_{r2}} - \frac{(\alpha_{12}^2 + \alpha_{23}^2)\frac{1}{\tau_a}}{\frac{1}{\tau_a^2}} = \frac{1}{\tau_a}$. This equation indicates that the coupling between the resonances helps drive the system meeting the critical coupling and hence enhance the absorption of the system.

**Spectrum Fitting and Q factor extraction.** We used a mathematical model to fit the spectrum obtained from the experiments and then extract the Q factor with the equation $Q = \dfrac{\omega_0}{2\gamma}$. Here, $\omega_0$ represents the resonance frequency and $\gamma$ denotes the damping rate of the resonance. The curve fitting equation is given by $A = \left| a + ib + \dfrac{C}{\omega - \omega_0 + i\gamma} \right|^2$, where a, b, and c are constant real numbers; $\omega$ is the angular frequency. The fitting is performed by "CFTool" toolbox in MATALB. In this paper, the R-square values of all curve fitting are larger than 92%.

# Acknowledgments

**Funding:** G.Z, I.H., T.A, M.T.U, and J.C.N, acknowledge support from Office of Naval Research Young Investigator Program (ONR YIP) Award (Grant number: N00014-24-1-2085) awarded to J. Ndukaife.

**Author contributions:** J.C.N. initiated and guided the project. G.Z and J.C.N. conceived the design. G.Z performed the numerical simulations and theoretical analysis, fabricated the metasurface and conducted the thermal emission experiments. I.H., M.T.U, and J.R.N contributed to the FTIR spectral measurements. T.A. contributed to fabricating the metasurface. M.H. contributed to optimizing the experimental set-up and band structure simulation. J.D.C contributed to the discussion of the results. All the authors contributed to editing the manuscript.

**Competing interests:** A patent application on the technology is pending.

**Data and materials availability:** Data underlying the results presented in this paper are not publicly available at this time but may be obtained from the authors upon reasonable request.

**Supplementary Materials**

**This PDF file includes:**
Supplementary Text
Figs. S1 to S6

# Supplementary Materials for

**Engineering thermal emission with enhanced emissivity and quality factor using bound states in the continuum and electromagnetically-induced absorption**

Guodong Zhu *et al.*

*Corresponding author. Email: justus.ndukaife@vanderbilt.edu

**This PDF file includes:**

Supplementary Text
Figs. S1 to S6

**Supplementary Text**

Thermal emission measurement experimental setup

The emission measurement is performed using a mercury cadmium telluride (MCT) detector on a Bruker VERTEX 70v Fourier transform infrared (FTIR) spectrometer, which is integrated with an in-house-built external rig for collecting and filtering the emission in the normal direction. As shown in Fig. S1, the sample is heated on a stage, and its emission is collimated using a 90° off-axis parabolic mirror. A pinhole is placed after the parabolic mirror to filter emission light that deviates from the normal direction. During the experiment, the pinhole is adjusted to the smallest possible size while maintaining a good signal-to-noise ratio, allowing that only a small angle (approximately 1°) of emission in the normal direction is collected. Then, another parabolic mirror is introduced to focus the collimated emission onto a second pinhole, which blocks the thermal emission outside the sample. Afterwards, the third parabolic mirror collimates the emission, followed by gold mirrors to direct the emission into the FTIR. Before the emission enters the FTIR port, a polarizer can be introduced to only allow the emission of one polarization to enter the FTIR.

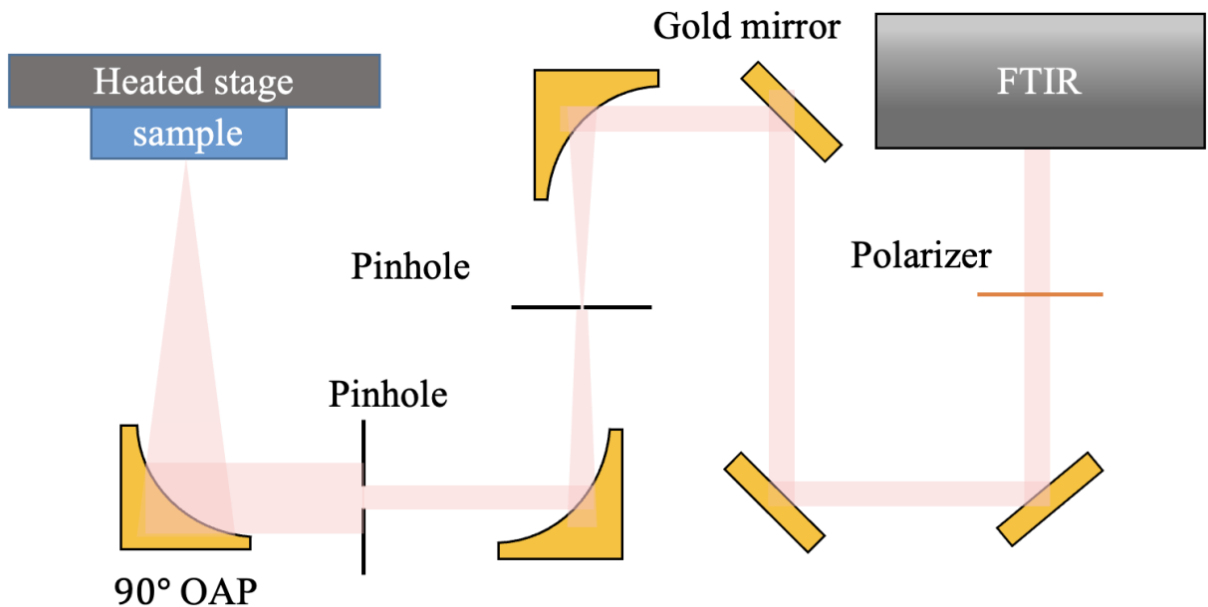

**Fig. S1.**

Top-view schematic of in-house thermal emission measurement set-up. OAP represents off-axis parabolic mirror.

Fabrication process

The fabrication process begins with the resistive physical vapor deposition (Angstrom Amod) of 10 nm chromium thin adhesion layer on top a clean glass substrate (2 cm × 2 cm), followed by electron-beam deposition of a 150 nm gold reflection layer and a 150 nm $Al_2O_3$ spacer layer. Then, a double layer of polymethyl methacrylate (PMMA) resist with different molecular weights (495K A4 and 950K A4) is spin-coated. The total thickness of the PMMA layer is approximately 400 nm, which is sufficient for creating patterns with a height of 100 nm. Subsequently, after the resistive physical vapor deposition of a 10 nm chromium conduction layer, a 4 mm × 4 mm pattern is transferred to the resist using 30 keV electron beam lithography (Raith eLiNE). The sample is then first dipped in the chromium etchant (Transene Chrome Mask Etchant 9030) to remove the conduction layer, followed by development in MIBK/IPA 1:3 (Kayaku Advanced Materials). Then, to fabricate gold ring nanostructures, an adhesion layer of 3 nm titanium and 100 nm of gold is vertically deposited via electron-beam physical vapor deposition (Angstrom Amod). Finally, the wet-chemical lift-off is performed in NMP 1165 remover (Microposit) on an 80 °C hot plate overnight to remove the PMMA.

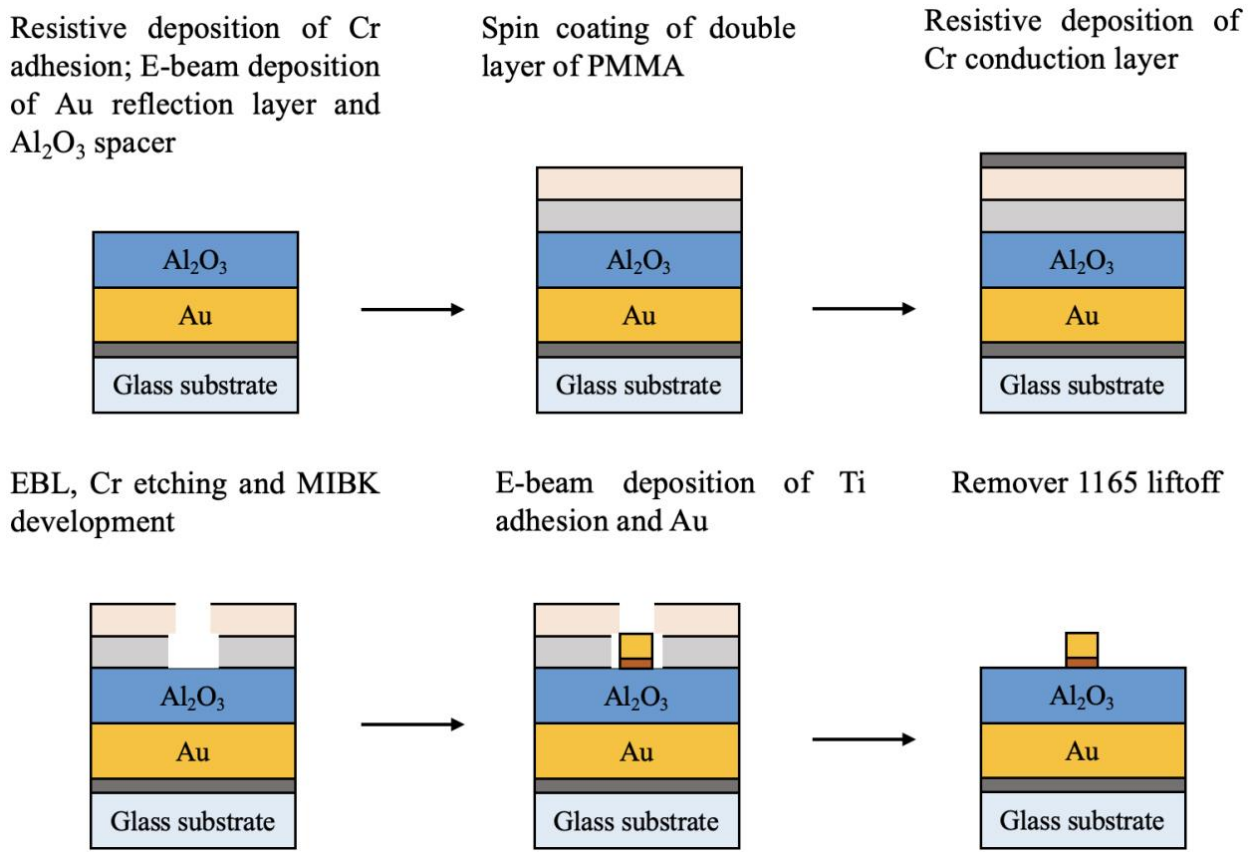

**Fig. S2.**

Fabrication flow of the thermal metasurface.

Spectrum tunability of the design

Fig. S3 shows that the emission spectrum is tunable in a wide spectrum range ((3 μm to 8 μm), which indicates a potential in various applications including sensing, communications, and radiation cooling.

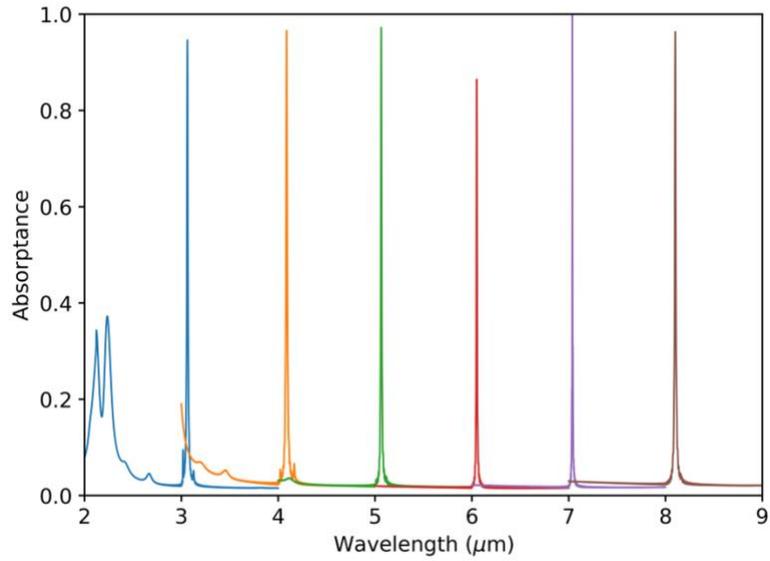

**Fig. S3.**
The resonance wavelength is tunable at a wide spectrum range.

Slight spectrum broadening and emissivity drop due to angular dispersion

By leveraging the high dispersion of the band, we achieve good spatial coherence and directional emission. However, measurements of the directional emission become more complicated. In the experiments, the response of the sample was collected over a range of angles centered around the normal direction. As shown in Fig. S1, a pinhole is used to filter emission light that deviates significantly from the normal direction. The emission light passing through the pinhole has an angular spread of approximately 1°. We assume that the emission evenly contributes to the signal within 0°-1° range. The overall spectrum is estimated using Lumerical simulation, as illustrated in Fig. S4. The overall spectrum exhibits slight spectral broadening and a drop in emissivity compared with 0° spectrum. As a result, the Q factor and emissivity of the experimental results are expected to be lower than those predicted by the simulation.

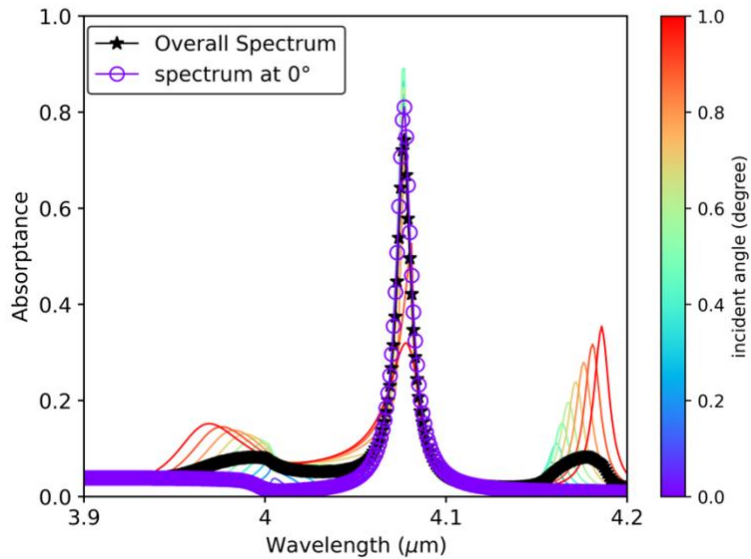

**Fig. S4.**

The overall spectrum of 1° angular spread. The colorful spectrum shows the absorption at various angles within 1°. The overall spectrum is the black line with start maker and the normal incident spectrum is the purple line with circle marker.

Characterize BIC for Resonance 1 and Resonance 3

To clarify that the Resonance 1 (M1) and Resonance 3 (M3) are associated with BICs, we perform a parameter sweep of the width ($w$) and angle (theta: $\theta$) of the segmented ring, as illustrated in Fig. S5. To simplify the interpretation of the results, we fix the height of the pillar at 300 nm and the thickness of spacer as 500nm because the three resonances are more separated under this condition. In Fig. S5(a), we keep the theta unchanged and tune the width of the segmented ring. We observe that as the width of the segmented ring decreases, the linewidths of M1 and M3 become narrower and eventually disappear. A similar behavior occurs when we tune the theta for M3 in Fig. S5(b), the linewidth vanishes when theta equals zero. However, for M1, there are two points at which the linewidth of M1 disappears. The first one occurs at the symmetry points ($\theta = 0°$), which indicates a symmetry-protected BIC (SP-BIC). The second one occurs at around $\theta = 100°$, which indicates a Friedrich-Wintgen BIC FW-BIC. In summary, we conclude that M1 is a dual-BIC resonance, combining both SP-BIC and FW-BIC characteristics, whereas M3 is a single SP-BIC resonance.

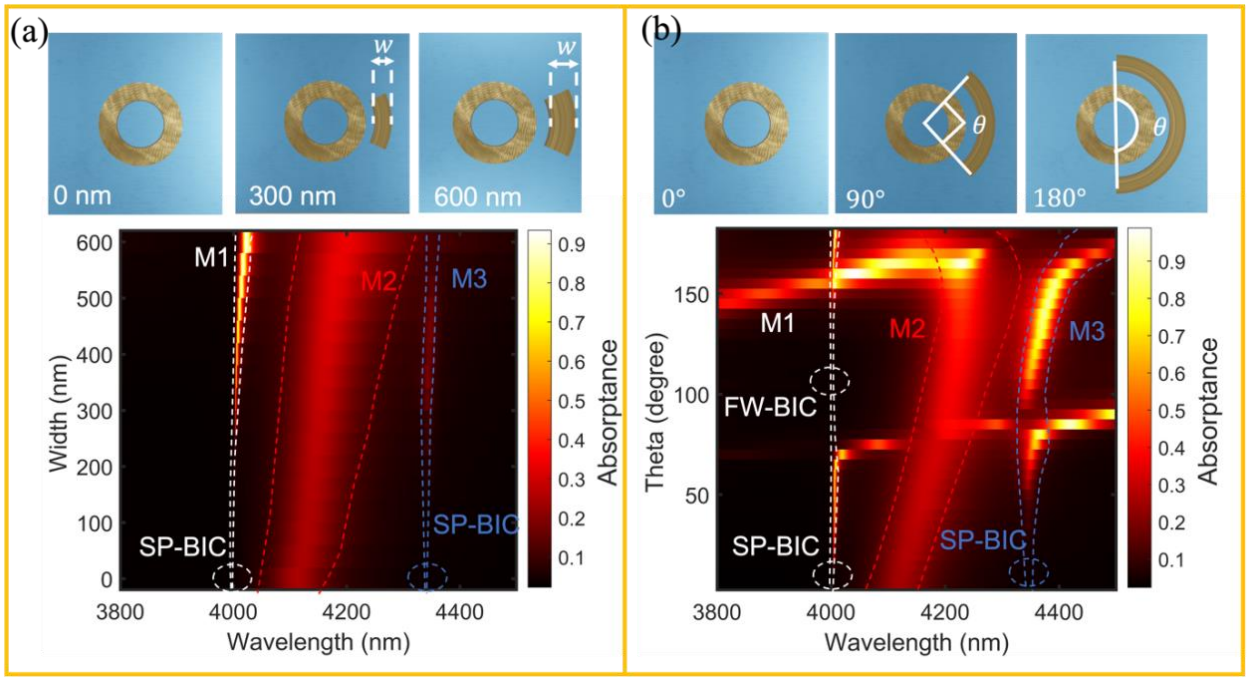

**Fig. S5.**

(a) and (b) shows the parameter sweep results refer to angle and width of the segmented ring for characterizing Mode 1 and Mode 3. In (a), we keep theta = 50° and sweep the width of the segmented ring form 0 nm to 600 nm. In (b), we keep the width = 290 nm and sweep the theta from 0° to 180°. From results, we can find the linewidth of Mod1 disappears at three conditions: theta = 0° or width = 0 (indicating SP-BIC) and theta around 100 (indicating FW-BIC). For Mode 3, the linewidth only disappears at theta = 0° or width = 0 (indicating SP-BIC).

Multipole decomposition

We employ an open-source MATLAB tool "MENP" to perform multipolar decomposition on Cartesian coordinates and compare the results for the three resonances under both off-optimal and optimal coupling conditions, as shown in Fig. S6. The scenario under off-optimal coupling is shown in Fig. S6(a): the bright state M2 exhibits broadband scattering with significant amplitude, while the dark states M1 and M3 are very weak due to weak coupling to free space. M2 is dominated by a magnetic dipole. The scenario under optimal coupling is illustrated in Fig. S6(b). The multipolar decomposition results show that M2 turns into a very sharp peak in the scattering spectrum while still maintaining the dominant magnetic dipole mode. The results demonstrate that no mode transition occurs during the coupling process.

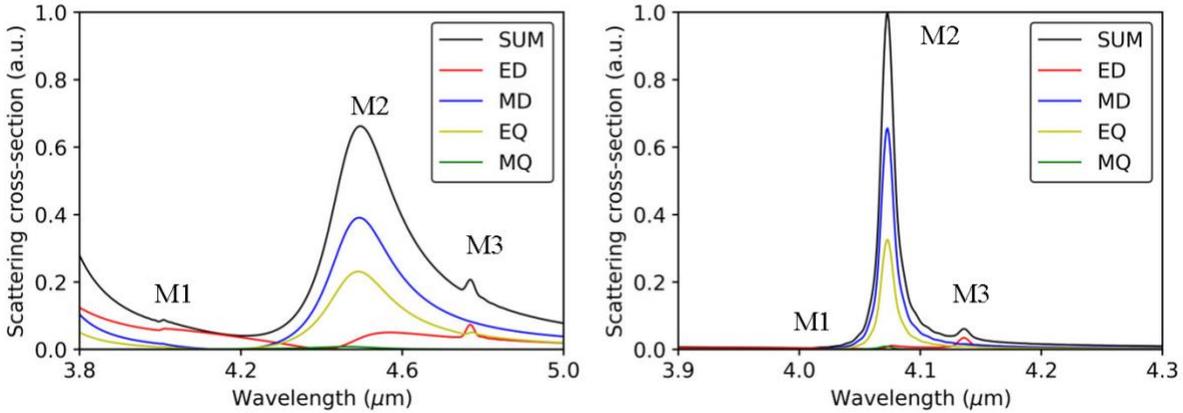

**Fig. S6.**

Multipolar decomposition results under both off-optimal and optimal coupling conditions. (a) shows the results under off-optimal coupling condition. For the structure parameters in this condition, we set the spacer layer thickness to 500 nm, the width of segmented ring to 200nm and the theta of segmented ring to 56 degrees. The other structure parameters are the same with optimal coupling condition. (b) shows results under optimal condition. The structure parameters are the same as parameters shown in the Fig. 1 in main manuscript.

6